# Deviations from the Summerfield scaling in the piezoelectric PbZr$_{0.57}$Ti$_{0.43}$O$_3$

M. Mahesh Kumar* and Z.-G. Ye

*Department of Chemistry, Simon Fraser University, Burnaby, BC, V5A 1S6 Canada*

Extending our earlier observations of the deviations from Summerfield scaling in a crystalline ferroelectric system of SrBi$_2$Ta$_2$O$_9$ (SBCT) [1], we have studied the scaling of conductivity in another ferro/piezoelectric perovskite crystalline oxide, PbZr$_{0.57}$T$_{0.43}$O$_3$ (PZT). The deviations from Summerfield scaling are found at T > T$_C$ (= 420 $^o$C). Correction to the Summerfield scaling is achieved by the application of the Random Barrier Model (RBM) and master cures are constructed with a value of the scaling function α ~3.2. As in the case of SBCT, deviations still occur at T > T$_C$. The master curves are constructed from the conductivity data σ'(*f*, T) fitted to a modified Josher's law in the frequency range 10- 10$^5$ Hz over the temperature region 30 – 480 $^o$C. The positive values of α indicate coulomb exchange between the interacting particles. This work provides an additional example which allows us to believe that crystalline systems as well may follow the scaling laws, not only amorphous glasses, as was previously believed.

Electrical properties of a large number of materials, eg., glasses, amorphous semiconductors, electron conducting polymers etc., are conspicuously common in that the general shape of the conductivity curves are very similar. In most of these solids, the dc conductivity follows the Arrhenius equations with universally similar ac conductivity. The shapes of the curves are so analogous that it is impossible to distinguish between the ionic and electronic conductivity. Such ac conductivity spectra of different temperatures can be scaled to fall on a single master curve indicating universality and a common underlying disorder.

In the absence of free charge carriers in disordered semiconductors, ac conductivity (σ') is different from dc conductivity (σ$_{dc}$) and σ$_{dc}$ increases as a function of temperature,[1] while hopping of ions results in the ac conductivity in ionic conductors, quantum mechanical tunneling gives rise to electronic conduction in disorder solids, even though the ac conductivity curves of these two different mechanisms look similar. Exceptionally, the common features among these two types of ac conduction are the "jumping and tunneling rates" and "the local mobilities", which allow us to scale different temperature curves into a single master curve.[2]

Taylor found the scaling behavior of in ionic conducting glasses by plotting the dielectric loss against scaled frequency.[3] The frequency axis was later modified by Isard using the product of frequency and dc resistivity given by,[4]

σ(ω)/ σ$_{dc}$ = F[C(ω/σ$_{dc}$)].            (1)

Time-temperature superposition principle (TTSP) is a correction to the Taylor-Isard scaling, whereby it is possible to calculate the value of the constant C.[2] TTSP describes the microscopic mechanisms under study in many electrical and mechanical systems. The spectral shapes of many of these systems are so similar that it is possible to force the data of different temperatures to fall on a single master curve, which indicates similarity in the physical mechanisms. This behavior also shows that the microscopic mechanism of a large number of system does not depend on temperature.[2]

A large number of systems such as inorganic solids and glasses, semiconductors and polymers have been found to follow TTSP.[5] However, one noticeable aspects of all these materials is that they are amorphous in nature. In crystalline disordered systems, such scaling behavior has not been known until recently, even though their conductivity spectra follow Jonscher's universal conduction laws. In a recent work, we have reported the scaling of the conductivity [σ'(*f*,T)] data of a Ca$^{2+}$ doped ferroelectric SrBi$_2$Ta$_2$O$_9$ (SBCT) to the modified TTSP by Summerfield equation and the corrections by using the random barrier model (RBM).[1] This was the first example of such scaling for a system other than amorphous solids. Deviations from the Summerfield scaling were found at T < T$_C$ (the ferroelectric Curie temperature) which were effectively corrected by the use of RBM model, at T >T$_C$ however, both models were still disobeyed. The σ'(*f*,T) data fits to a modified Jonscher expression, indicating multiple relaxations, which were unknown in crystalline ordered systems, as opposed to amorphous glasses.[1]

A question that naturally arises from the above investigations is whether such a scaling and related phenomena are limited to only one particular crystalline system or they are applicable to other systems in general. To answer this question we have applied the scaling laws to other crystalline systems including those that are ferroelectrically ordered. In this paper, we demonstrate the evidence of the adherence of scaling laws by the prototypical ferro/piezoelectric PbZr$_{0.57}$Ti$_{0.43}$O$_3$ (PZT) by fitting the σ'(*f*, T) data to Summerfield and RBM scaling laws. PZT is a well known material which is used extensively in ceramic form in actuator and transducer applications. our results show that interestingly, not only do the σ'(*f*,T) data scale to the corrections to Summerfield scaling, but also follow the modified Jonscher's expression that was proposed for SBCT in our earlier report.[1] This suggests that such scaling law may be universally applicable to those materials in which chemical disorder on a certain crystallographic site being a common feature.

Ceramic samples of $PbZr_{0.57}Ti_{0.43}O_3$, i.e. with composition within the morphotropic phase boundary (MPB) were prepared by solid state reactions starting from stoichiometric amounts of PbO, $ZrO_2$, and $TiO_2$. The green mixture was ground, pressed into pellets, and heated to 950 °C for 2 h. The calcined pellets were crushed and ball milled with $ZrO_2$ balls for 24 h, before a 5% (by weight) PVA binder was added to the mixture. Final sintering was carried out at 1200 °C for 2 hrs on pressed pellets (13 mm in dia., 1.5 mm thick). X-ray diffraction patterns confirmed the formation of a pure perovskite single PZT phase and scanning electron microscopy showed with a grain size of ~3μm. Large sides of the pellet were deposited with Au as electrodes. Complex impedance $[Z^*(f, T)]$ was measured in the temperature range 30- 500 °C with the frequency varying between $1 - 10^5$ Hz, using a Solartron 1260 impedance analyzer interfaced to a 1290 dielectric interface. Ac conductivity was calculated using the expression,

$$\sigma' = Y'_* (t/A) = [Z'/(Z'^2 + Z''^2)]_* (t/A), \quad (2)$$

where $Z'$ and $Z''$, are the real and imaginary parts of impedance, $Y'$ is the real part of the admittance, t the thickness (m) and A the area ($m^2$) of the sample, respectively. Temperature was stabilized (± 0.5 °C) at every 10 °C intervals, while sweeping $Z^*$ as a function of frequency, $f$.

The ac conductivity ($\sigma'$) derived from Eq. (2) is shown in Fig. 1, as a function of $f$, at temperatures $30 \leq T \leq 460$ °C. The low frequency $\sigma'(f, T)$ data show little or no variation, corresponding to the frequency-independent part, $\sigma_{dc}$. At higher frequencies, the $\sigma'(f, T)$ spectrum shows a dispersion, indicating the onset of ac conductivity. At $f > 10^5$ Hz the spectra of different temperatures tend to merge onto a single curve, which is commonly seen in conventional disordered solids. At high temperatures, there is a marked decrease in the frequency dispersion of $\sigma'(f, T)$, with dc conductivity dominating over the frequency range studied.

Ac conductivity, $\sigma'(f, T)$, is generally described using Josher's power laws, which were found to be universal in many disordered solids, in particular ionic conducting glasses, heavily doped ionic crystals and many polymers.[6] Recently they were also observed in crystalline solids of ferroelecrics.[7] In a number of cases, in addition to the Jonscher's power laws, another term exists that is known as near-constant loss (NCL). NCL shows up as a straight line on a log-log plot of conductivity vs. frequency, in the high frequency range of $\sigma'(f, T)$ data. Thus, a complete expression consists of Jonscher's term plus an NCL term, as given by,[8]

$$\sigma'(f) = \sigma_{dc}[1+(\omega/\omega_p)^n] + A.\omega, \quad (3)$$

where $\omega (=2\pi f)$ and $\omega_p (=2\pi f_p)$ are the angular measuring and relaxation frequencies, respectively; $\omega_p$ signifies the cross-over from dc to ac conductivity regime; A is a constant, which is temperature independent at high temperatures in the NCL part.

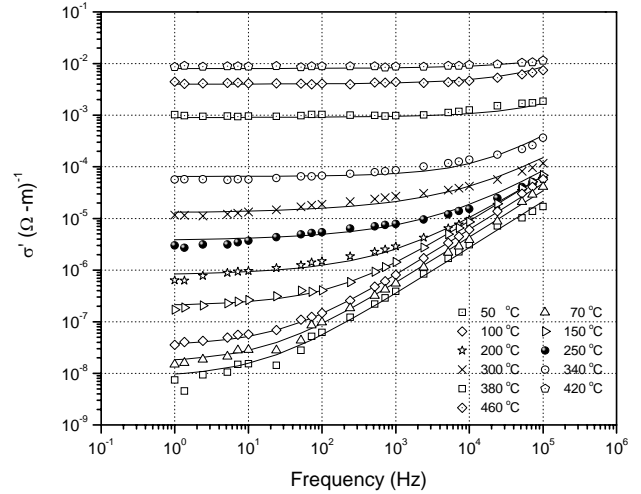

FIG. 1. Ac conductivity as a function of frequency at different temperatures of $PbZr_{0.57}Ti_{0.43}O_3$ ceramic. Lines are fits to Eq. (4).

The non-linear behavior at high frequencies did not necessitate the inclusion of NCL term during the fitting of the $\sigma'(f, T)$ data to Eq. (3). However, the Jonscher's term alone did not yield a completely positive fit and deviations were predominant in the regions of the crossover from dc to ac conductivity. To overcome this deficiency, a modified Jonscher's expression was used with multiple relaxations:

$$\sigma'(f) = \sigma_{dc}[1+(f/f_p)^n + (f/f_q)^m], \quad (4)$$

where $f_p$ and $f_q$ are the relaxation frequencies, and $n$ and $m$ are exponent whose values generally lie between 0 and 1. A fit of the ac conductivity to the above equation yields a satisfactory fit, as shown by the solid lines in Fig. 1, with the values of $n$ and $m$ being 0.4 and 0.9 respectively. The value of $n$, in particular is close to the NCL. We previously used this modified Jonscher's expression to fit the $\sigma'(f, T)$ data of SBCT, in which the values of $n$ =0.4 and $m$=0.7 were obtained.[1]

As clearly seen from Fig. 1, $\sigma_{dc}$ is almost temperature independent at $f < f_p$. With increasing frequency ($f > f_p$) conductivity rises slowly, as opposed to a steep rise in materials with a single relaxation. This slow increase indicates the presence of multiple relaxations, which necessitated the addition of an incremental term for an additional relaxation frequency ($f_q$) in Eq. (3) to achieve perfect fitting. At high temperatures (T > 340 °C), $f_q$ is temperature independent and $\sigma'(f, T)$ fits to a normal Jonscher's expression.

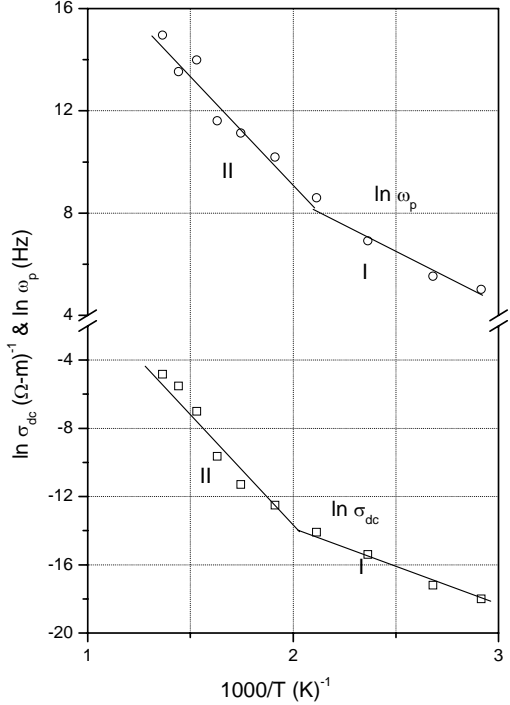

FIG. 2. Arrhenius fits of dc conductivity and the relaxation frequency $\omega_p$. The two different regions of conductivity are indicated as I and II.

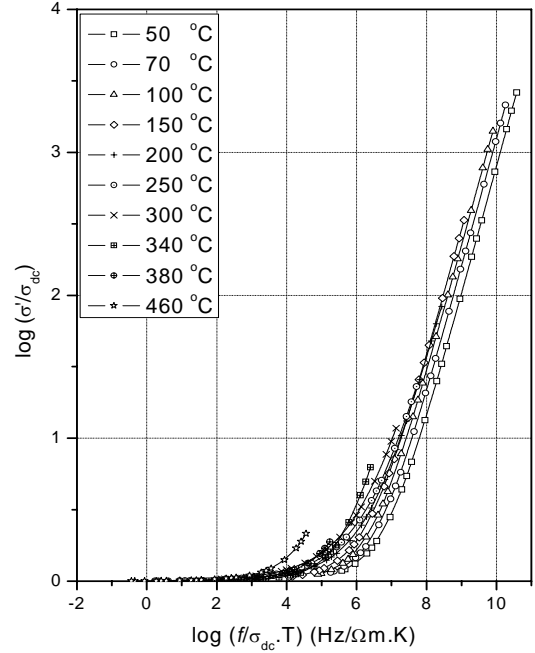

FIG. 3. Master curves of conductivity plotted at different temperatures. A deviation from Eq. (6) occurs as a function of temperature, overwhelmingly at $T > T_c$ and marginally at $T < T_c$.

The different exponents $n$ and $m$ indicate different conduction mechanisms in disorder solids. In a recent tabulation of the relaxation times of various materials, Sidebottom found that the exponents are related to the dimensionality of conduction pathways.[9] Different values of exponents were associated with various conduction mechanisms in the complete frequency range of studies. The exponent $n = 0.4$ ($f > f_p$) is in line with the exponent reported in the superionic conductor $K_{2x}Mg_xTi_{8-x}O_{16}$. Sidebottom predicts that a value 0.4 (±0.1) pertains to a one-dimensional (1D) system of ionic conduction, as exemplified by $K_{2x}Mg_xTi_{8-x}O_{16}$ with hollandite structure,[10] whereas, the exponent $m = 0.9$ ($f > f_q$) is close to the NCL value of 1, indicating motion of mobile charges and ionic species in asymmetric double well potentials.[11] An exponent value equivalent to 0.9 was recently observed in alkaline germanate glasses.[12]

In the Jonshcer's expression of Eq. (4), $\sigma_{dc}$ and $f_p$ are temperature-assisted hopping processes and follow Arrhenius laws with equal activation energies. Fig. 2 shows the logarithmic conductivity and relaxation frequency $f_p$ as a function of reciprocal temperature. The conductivity curve could not be fitted to a single Arrhenius relation, as it is made up of at least two such thermally activated processes. Similarly $f_p$ also follows two processes with two activation energies. The calculated activation energies and the temperature ranges of fitting are tabulated in Table 1. The activation energies of $f_p$ are almost half of those observed for $\sigma_{dc}$. This near halving of $E_a$ for $f_p$ is unexpected and is contrary to the conventional view that the $E_a$s of $\sigma_{dc}$ and $f_p$ are generally equal. Another important observation is that $f_p$ does not deviate around $T_C$ (= 420 °C), as seen in SBCT,[1] but a deviation into another Arrehenius regime at a low temperature (~200 °C). As revealed from the paramagnetic resonance experiments, a value of $E_a > 1$ eV in perovskite relates to the motion of the oxygen vacancies, whereas the low values of activation energy in $\sigma_{dc}$ results from mixed oxygen and ionic conductivity at low temperatures.[13]

In the ensuing, we will discuss the adherence of σ'($f$, T) of PZT to scaling laws. The temperature dependent conductivity can be superimposed onto a single master curve by Taylor.[3] The mastercurve also indicates the independence of the curve shape on varying temperatures. The similarity in the present shapes of the σ'($f$, T) curves allow us to scale them to the TTSP, modified by Summerfield by replacing the characteristic frequency $\nu_0$ with $\sigma_{dc}T$ in the scaling function,

$$\sigma(\omega)/\sigma_{dc} = F(f/f_0). \quad (5)$$

Table 1.

| Temperature range (°C) | Activation energies $E_a$ (eV) calculated from | |
|---|---|---|
| | $\sigma_{dc}$ | $\omega_p$ |
| 30-200 (I) | 0.45 | 0.21 |
| 200-480 (II) | 1.47 | 0.74 |

Thus the Summerfield scaling is given by,[14]

$$\sigma(\omega)/\sigma_{dc} = F(f/\sigma_{dc}.T). \qquad (6)$$

Figure 3 shows the master curves obtained by scaling of Eq. (6) to σ'(*f*, T). A certain deviation from the Summerfield scaling appears at low and high frequencies. Even though the high temperature curves appear to converge on a master curve, the deviation at high frequencies indicates that the shape of the σ'(*f*, T) is not completely independent of temperature. In particular, the low temperature (50 -200 °C) curves are located far away from the high temperature curves, suggesting the temperature dependence of the charge mobility and dynamics. It is important to note that the complete deviation of the 460 °C (> $T_C$) curve from the high temperature curves. This phenomena was similar to what we observed in ferroelectric SBCT.[1] The failure of Summerfield scaling is evidenced by the lack of a coherent master curve, which also indicates that the dynamics of conductivity are not the same in these frequency and temperature ranges of investigation. Such deviations from Summerfield scaling were previously seen in alkaline tellurite glasses at high frequencies, by Murugavel *et al.*,[15] who explained that the 'structural peculaliarities' in the materials could result in different conduction pathways, giving rise to the deviation from the Summerfield scaling. In the ferroelectric perovksite oxides such as PZT, the ferroelectric to paraelectric transition is always associated with a structural transformation from a non-centrosymmetric to a centro-symmetric phase, which could induce a scaling deviation as seen at T > $T_C$. Even at T < $T_C$ there is a gradual structural change leading to a first order phase transformation ultimately resulting in a complete structural change. As indicated by $\sigma_{dc}$ vs T, the conductivity mechanism, is not the same in the temperature ranges measured. The considerable scaling deviations seen at low temperatures are due to the effect of a differing conduction mechanism compared to the high temperature. The concurring spectral shapes at low temperatures, although indicate similarities in conduction mechanism, are temperature dependent.

The corrections to the Summerfield scaling were proposed in the random barrier model (RBM), which accounts for the inhomogenieties created by the chemical composition. One of the corrections was provided by Baranovskii *et al.*[16] by introducing a scaling function α, which indicates the deviation from the Summerfield scaling. Through the simulation of a single particle using the percolation approach, a formula for the Summerfield correction was given as follows,

$$\sigma(\omega)/\sigma(0) = G[(f/\sigma_{dc}T).T^{\alpha}]. \qquad (7)$$

The value of α indicates type the of interaction between the particles. Generally, α takes negative values for the conductivity arising from non-interacting particles.[17]

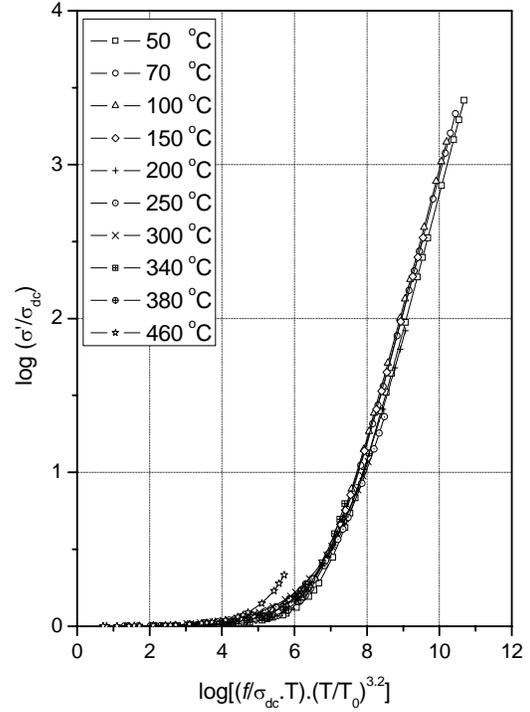

FIG. 4. Master curves of conductivity isotherms plotted at different temperatures using Eq. (7). A clear improvement in scaling at T < TC has been achieved.

Figure 4 shows the data of σ'(*f*, T) fit to Eq. (6), with a positive value of α = 3.2. With $T_0$ being fixed at 300 K, it is clear that the scaling has been improved dramatically, with all the curves of T ≤ 380 °C falling on to a single master curve. However, deviations at T > $T_C$ (~ 420 °C) can still be observed, with other curves below $T_C$ falling on to the master curve. Therefore, eventhough Eq. (6) is a better scaling function, it still fails at T > $T_C$. In contrast, scaling of the conductivity was observed in amorphous glasses even at temperatures greater than their glass transition temperatures. The reasons for the failure at T > $T_C$ in PZT and SBCT may be the associated with the ferro to paraelectric structural transformations occurring at $T_C$. The above analysis using Summerfield [Eq. (6)] and RBM scaling [Eq. (7)] indicates a possibility that crystalline systems also may follow the scaling laws enunciated for mostly amorphous solids..

The values of α obtained for PZT are even higher than those in amorphous glass systems. The maximum value obtained in alkali tellurite glasses for low concentrations of sodium oxide addition was 2.5 and a positive value points to a mechanism of interacting particles.[15] With increasing interaction of the particles, it is believed that the value of α increases in the purview of the RBM theory. However, increasing interactions between the particles does not seem to affect the scaling of the spectra and the failure of the Summerfield scaling in this regard can be attributed to the structural peculiarities, as was also the case for alkali tellurite glasses.

In conclusion, we have shown in the above investigation the applicability of scaling laws to another very important ferroelectric crystalline system of PbZr$_{0.57}$Ti$_{0.43}$O$_3$, providing an additional example of materials other than amorphous glasses. The failure of Summerfield scaling could be attributed to the structural asymmetries, which create inhomogeneous potential environments with restricted ion movement pathways due to the compositional disorder at the B-site. The RBM theory was applied to correct this anomaly to obtain a single master curve for the σ'($f$, T) data at different temperatures. However, both theories fail at T > T$_C$ as also seen in the other ferroelectric, SBCT. More systems are needed to be tested to claim universality of these scaling laws in crystalline ordered systems in general.

The authors thank Natural Sciences and Engineering Research Council of Canada (NSERC) for financial support and J. Cheng for the help in sample preparation.

**References**

*Corresponding author
Now at: Institute for Chemical Process and Environmental Technology, National Research Council of Canada, Ottawa, Canada, K1A 0R6. e-mail: mahesh.matam@nrc.ca